\documentclass[12pt]{article}
\usepackage{amssymb}

\begin{document}




{\hbox to\hsize{\hfill June 2015 }}

\bigskip \vspace{3\baselineskip}

\begin{center}

{\bf \Large 

Quantum relaxation of the Higgs mass }

\vskip 0.4cm

\bigskip

\bigskip

\bigskip

\bigskip

{\bf Archil Kobakhidze  \\}

\smallskip

\bigskip

{ \small \it

ARC Center of Excellence for Particle Physics at the Terascale, \\
School of Physics, The University of Sydney, NSW 2006, Australia \\
E-mail: archil.kobakhidze@sydney.edu.au
\\}

\bigskip

\bigskip

\bigskip

\bigskip

\bigskip

{\large \bf Abstract}

\end{center}

\noindent
{\small I put forward a qualitatively new dynamical mechanism for solving the electroweak hierarchy problem that does not require new physics at the electroweak. I argue that the infrared fluctuations of the gravitational field may provide a partial screening of the Higgs mass, similar to the infrared screening of the electric charge in quantum electrodynamics.}

\bigskip

\bigskip



The discovery of the 125 GeV Higgs boson with properties consistent with an elementary scalar, together with non-observation of any sign of new physics at the LHC so far, makes a search for qualitatively novel realisations of the naturalness paradigm paramount. One interesting theoretical possibility has been recently proposed in \cite{Graham:2015cka}. According to this proposal the Higgs mass is driven to a value much smaller than the ultraviolet cut-off [being as large as $\sim 10^8$ GeV] by a peculiar Higgs-axion dynamics in the early inflationary universe.  

I would like to propose another unorthodox solution to the mass hierarchy problem, which is inspired by Polyakov's idea on the infrared screening of the cosmological constant \cite{Polyakov:1982ug} (see also \cite{Jackiw:2005yc}).  The purpose of this brief note is to demonstrate the key aspect of the mechanism, without neither going into much phenomenological details, nor attempting at ultraviolet completion to establish a fully consistent theory. To this end, I consider an effective theory of the electroweak Higgs doublet $\Phi$ coupled to gravity valid at some high-energy scale $\mu_{\rm UV}$.  The effective  action reads:
\begin{equation}
S_{\mu_{\rm UV}}=\int d^4x \sqrt{-g}\left[\frac{R(g)}{16\pi G_{\rm N}} + \partial_{\mu}\Phi^{\dagger} \partial^{\mu}\Phi -V(\Phi^{\dagger}\Phi)+...\right]~,
\label{1}
\end{equation} 
where, 
\begin{equation}
V(\Phi^{\dagger}\Phi) = \Lambda +M_{\Phi}^2\Phi^{\dagger}\Phi+\frac{\lambda}{2}\left(\Phi^{\dagger}\Phi\right)^2~,
\label{2}
\end{equation} 
is the Higgs potential with a cosmological constant $\Lambda$ and the ellipsis meant to contain all the remaining terms, including  high-dimensional operators which are allowed by symmetries of the theory.  These terms are irrelevant for the sake of the argument and I ignore them.  
All the tree-level parameters in the above  action ($\Lambda, M_{\Phi}, \lambda,...$) are defined at the cut-off scale $\mu_{\rm UV}$. This implicitly assumes that they have been renormalised within a more fundamental theory valid beyond $\mu_{\rm UV}$. Therefore, 
 one naturally expects that $|M_{\Phi}|\sim |\Lambda|^{1/4}\sim \mu_{\rm UV}$ at the cut-off scale.\footnote{This is not necessarily true. If certain symmetries, such as supersymmetry  or scale invariance, are enforced upon the effective low-energy theory from its UV completion, the effective Higgs mass parameter $M_{\Phi}$ can be smaller than the ultraviolet cut-off \cite{Kobakhidze:2014afa}. This is the standard paradigm.} In other words,  the model is not tuned. 

The theory  (\ref{1}) is nonrenormalisable and this complicates the discussion within the full general relativistic framework. Instead, following  \cite{Polyakov:1982ug}, I restrict the full metric $g_{\mu\nu}$ to be a conformally flat metric:
\begin{equation}
g_{\mu\nu}=\phi^2\eta_{\mu\nu}~,
\label{3}
\end{equation} 
where $\eta_{\mu\nu}$ is just the flat Minkowski metric. For conformally flat spacetime (\ref{3}) one computes:
\begin{eqnarray}
R(g)=-6\phi^{-3}\Box \phi~,~~\sqrt{-g}=\phi^4~,
\label{4}
\end{eqnarray}
where $\Box$ is the flat spacetime D'Alambertian operator.  It is convenient to introduce a new set of fields obtained from the old one  by the following field redefinitions:
\begin{eqnarray}
S=\sqrt{6}M_P\phi~,~~H=\phi \Phi~,
\label{5}
\end{eqnarray}
where $M_P=1/\sqrt{8\pi G_N}\approx 2.4\cdot 10^{18}$ GeV is the reduced Planck mass. The action (\ref{1}) then can be rewritten as: 
\begin{eqnarray}
S_{\mu_{\rm UV}}=\int d^4x \left[\frac{1}{2}\partial_{\mu}S\partial^{\mu}S +{\cal D}_{\mu}H^{\dagger}{\cal D}^{\mu}H
-\frac{\lambda}{2}(H^{\dagger}H)^2-\frac{\lambda_{\rm M}}{2} S^2H^{\dagger}H-\frac{\lambda_{\rm \Lambda}}{4}S^4 +...\right] ~,
\label{6}
\end{eqnarray} 
where ${\cal D}_{\mu}=\partial_{\mu}-\frac{1}{S}\partial_{\mu}S$. In Eq. (\ref{6}), and in what follows, the Lorentz indices are contracted using the Minkowski metric. Note that the Higgs mass term and the cosmological constant turn into marginal operators and the corresponding parameters  become dimensionless:
\begin{eqnarray}
\lambda_{\rm M}(\mu_{\rm UV})=\frac{M_{\Phi}^2}{3M_P^2}~, 
\label{7}
\\
\lambda_{\rm \Lambda}(\mu_{\rm UV})= \frac{\Lambda}{9M_P^4}~. 
\label{8}
\end{eqnarray} 
Hence, the action (\ref{6}) exhibits a manifest classical scale invariance, $x^{\mu}\to \xi x^{\mu},~S\to \xi^{-1}S,~H\to \xi^{-1}H$, which is softly broken upon quantisation by the logarithmic dependence of the dimensionless couplings on the renormalisation scale $\mu$ ($\mu<\mu_{\rm UV}$). In particular, the renormalisation group (RG) running of (\ref{7}) and (\ref{8}) are determined, respectively, by the following one-loop RG $\beta$-functions:
\begin{eqnarray}
\beta_{\lambda_{\rm M}}=\frac{\lambda_{\rm M}}{16\pi^2}\left(\lambda_{\rm M}+\lambda_{\Lambda}+3\lambda\right)~, 
\label{9}
\\
\beta_{\lambda_{\Lambda}}= \frac{1}{64\pi^2}\left(\lambda_{\rm M}^2+2.5\lambda_{\Lambda}^2\right)~, 
\label{10}
\end{eqnarray} 
where we have omitted contributions from Higgs-Yukawa and gauge interactions for simplicity. Note, that (\ref{9}) exhibits fixed-point of the RG flow at $\lambda_{\rm M}=0$, which corresponds to the natural decoupling limit between gravity ($S$) and the Higgs sectors \cite{Foot:2013hna}. For $\beta_{\Lambda_{\rm M}}>0$ this is an infrared fixed-point. Also, according to Eq. (\ref{10}),  $\lambda_{\rm M}=0$ fixed point  triggers $\lambda_{\Lambda}\approx 0$ (quasi) fixed-point. This RG behaviour is at the heart of the infrared quantum relaxation of the cosmological constant \cite{Polyakov:1982ug} and the electroweak scale simultaneously. More specifically, assuming $\lambda_{\rm M}>>\lambda_{\Lambda}, \lambda$, we can easily solve the RG equation for $\lambda_{\rm M}$. Setting the infrared scale at the Higgs mass, $\mu=M_{\rm H}$, we find: 
\begin{equation}
\lambda_{\rm M}(M_{\rm H})\approx \frac{16\pi^2}{\ln\left(\frac{M_{\rm H}}{\mu_{\rm UV}}\right)}~.
\label{11}
\end{equation}      
To obtain the desired Planck/weak scale hierarchy one must have $\lambda_{\rm M}(M_{\rm H})\sim 10^{-32}$, which for the above perturbative result imply $\mu_{\rm UV}>>M_{\rm P}$. It could be though that in these regime the leading log perturbative calculations must be improved by employing non-perturbative techniques \cite{Jackiw:2005yc}. Also,  although I have no proof, but it seems unlikely that the hierarchy generated in (\ref{11}) can be undone through the fluctuations of the remaining  unimodular part of the metric tensor \cite{Polyakov:1982ug}. 

As an alternative model, one can consider a possibility for gravity to propagate in extra spatial dimensions. In that case the gravitational fluctuations will induce faster, power-law running of $\lambda_{\rm M}$ (as oppose to logarithmic running in 4 dimensions) and hence the desired hierarchy will be generated for smaller values of the ultraviolet cut-off \cite{Chaichian:1999ei}.  

To conclude, I demonstrate (at least qualitatively) that the electroweak mass hierarchy may originate due to the infrared quantum fluctuations of the  gravitational field which triggers quantum relaxation of the Higgs mass and perhaps the cosmological constant as well. It would be interesting to study this proposal in more quantitative manner.

\subsection*{Acknowledgements}
A part of this work was carried out  while I was attending the Gordon Research Conference on Particle Physics at the HKUST Jockey Club Institute for Advanced Study.  I would like to thank the organisers and participants for creating an enjoyable and stimulating atmosphere during the conference. I also would like extend my gratitude to Prof Jin Min Yang for his warm hospitality at the Kavli Institute for Theoretical Physics China, where the paper has been finalised. The work was supported in part by the Australian Research Council and also by the Rustaveli Science Foundation under the grants No. DI/8/6-100/12 and No. DI/12/6-200/13.

\end{document}